\documentclass[useAMS,usenatbib]{mn2e}
\usepackage{graphicx}
\usepackage{times}
\title[Ho II X-1 : the case against a 1000~M$_{\odot}$ black hole.]{A deep {\it XMM-Newton\/} observation of the ultraluminous X-ray source Ho II X-1 : the case against a 1000M$_{\odot}$ black hole.}
\author[M.R. Goad, T.P. Roberts, J.N. Reeves, P. Uttley]{
M.R. Goad$^{1}$\thanks{E-mail: mrg@star.le.ac.uk},
T.P. Roberts$^{1}$, J.N. Reeves$^{2,3}$, and P. Uttley$^{2}$\\ $^{1}$X-ray
\& Observational Astronomy Group, Dept. of Physics and Astronomy,
University of Leicester, University Road, Leicester, LE1 7RH, UK\\
$^{2}$Exploration of the Universe Division, NASA Goddard Space Flight Center, Greenbelt Road, Greenbelt, MD 20771, USA\\$^{3}$Department of Physics \& Astronomy, Johns Hopkins University, 3400 N Charles Street, Baltimore, MD 21218, USA}
\begin{document}

\newcommand{\mdot}{M$_{\odot}$}

\date{Accepted 200x June xx. Received 2005 June xx; in original form 2005 June
  xx}

\pagerange{\pageref{firstpage}--\pageref{lastpage}} \pubyear{2005}

\maketitle

\label{firstpage}

\begin{abstract}
We present results from a 112~ks long look by {\it XMM-Newton\/} at
the ultraluminous X-ray source (ULX) Holmberg II~X-1, long thought to
be the one of best candidates for the missing class of intermediate
mass black holes (IMBHs). Our data comprises the first high quality
{\it XMM-Newton\/}/RGS spectrum of an ULX, and an {\it
XMM-Newton\/}/EPIC spectrum with unprecedented signal-to-noise.  A
detailed timing analysis shows that any variability on time-scales of
minutes to hours is very weak ($<$ few per cent fractional rms),
though larger amplitude variations on much shorter time-scales could
be hidden by photon counting statistics.  This result suggests that if
Ho II X-1 harbours an IMBH, then we are observing this source in a
highly unusual and atypical state when compared with the known
variability behaviour of other accreting systems of large
mass. Moreover unlike Galactic X-ray binaries, our spectral analysis
indicates the possible presence of an optically-thick low temperature
corona.  Taken together our timing and spectral analysis suggests that
the compact companion is most-likely a high luminosity analogue of
black hole binary systems similar to GRS~1915+105, the Galactic
microquasar, harbouring a compact object of mass no greater than $100$
\mdot.

\end{abstract}


\section{Introduction}
{\it Einstein} and {\it Rosat} observations of nearby galaxies
revealed a new population of point-like extra-nuclear X-ray sources
with luminosities far in excess of the Eddington limit for spherical
accretion onto a 10 \mdot\ black hole ($L_{x} > 10^{39}$ erg~s$^{-1}$,
e.g. Fabbiano \& Trinchieri 1987; Roberts \& Warwick 2000), the
so-called ultraluminous X-ray sources (ULXs).  In fact, the
$\sim10^{39}-10^{41}$~erg~s$^{-1}$ X-ray luminosities of ULXs, if due
to sub-Eddington rate spherical accretion onto a compact object, imply
masses of $\sim 10^{2}$--$10^{4}$~\mdot.  ULXs may therefore represent
the so-called ``missing class'' of intermediate-mass black holes
(hereafter IMBHs), the purported primordial building blocks of
super-massive black holes and galaxies (e.g. Ebisuzaki et al. 2001;
Madau \& Rees 2001).  The best current evidence in favour of IMBHs
underlying ULXs derives from recent {\it XMM-Newton\/} spectroscopy of
several ULXs, where a ``soft excess'' in the spectra may be fit by a
relatively low temperature ($\sim 150$ eV) accretion disc model
(e.g. Miller et al. 2003; Miller, Fabian \& Miller 2004).  This is
considerably cooler than found in stellar-mass black hole X-ray
binaries (XRBs), and points to a larger, cooler disc surrounding a
more massive ($\sim 10^{3}$\mdot) black hole.  However, X-ray spectral
results based on continuum modelling may be open to some ambiguity, so
these detections are not necessarily conclusive in all cases (Roberts
et al. 2005).  Other possibilities involving stellar-mass black holes
have been mooted to explain the extreme X-ray luminosities of ULXs,
notably anisotropic radiation and/or relativistic beaming (King et
al. 2001; K{\"o}rding et al. 2002) or truly super-Eddington accretion
discs (Begelman 2002; Ebisawa et al. 2003).

A more direct determination of the mass of the compact accretor in
ULXs is the key to further progress in this field.  Perhaps the best
method for determining the mass of the compact object in ULXs from
X-ray data is through a comparison of their Power Spectral Density
(PSD) with those of XRBs and Active Galactic Nuclei (AGN). The PSD of
all known accreting systems are remarkably similar (Edelson \& Nandra
1999, Uttley et al. 2002).  In AGN, PSD shapes corresponding to both
low `hard' accretion state systems (NGC~3783, Markowitz et al. 2003)
and high `soft' accretion state systems (NGC~4051, M$^{\rm c}$Hardy et
al. 2004) previously seen in XRBs are now seen.  Significantly, the
{\it location\/} of the characteristic PSD {\it break-frequency\/}
scales inversely with the mass of the black-hole (Uttley et al. 2002,
Markowitz et~al. 2003, Vaughan et al. 2005) for objects with masses
ranging from 10 \mdot for a typical stellar mass black hole (eg. Cyg
X-1, $\nu_{\rm break}\sim$1~Hz) to $10^{5}$ -- $10^{8}$\mdot\ for the
central supermassive black hole in nearby Seyfert~1 galaxies. For the
latter, black hole masses have been independently verified using both
the stellar velocity dispersion and reverberation mapping techniques
(Gebhardt et al. 2000, Ferrarese et~al. 2001).  This method has been
applied to ULX data on at least two previous occasions (Cropper et
al. 2004, Soria et al. 2004); though putative breaks were detected,
the data were insufficient to place strong constraints on the black
hole mass in either case.

In order to progress, one requires the best possible X-ray data from
an ULX.  Here we present an X-ray spectral and timing analysis for an
unprecedented long look with {\it XMM-Newton\/} at the ULX
Holmberg~II~X-1 (hereafter Ho II X-1).  This is one of the most
luminous ($L_X\sim 1 - 2 \times 10^{40}\rm ~erg~s^{-1}$) nearby ULXs,
located in the $d \sim 3.2$ Mpc dwarf star-forming galaxy Holmberg~II.
Of the 87 candidate ULXs catalogued by Colbert \& Ptak (2002), Ho
II~X-1 had the highest observed {\it ROSAT\/} count rate making it an
ideal target for PSD analysis.  It is a very well-studied object; {\it
ROSAT\/} and {\it ASCA\/} observations (Zezas, Georgantopoulos \& Ward
1999; Miyaji, Lehmann \& Hasinger 2001) revealed a relatively soft
X-ray spectrum, and variability consistent with an accreting black
hole.  The detection of a soft excess in the X-ray spectrum, that can
be modelled as the accretion disc around a $\sim 10^3$\mdot IMBH, was
confirmed in three short ($\sim 10$ ks) {\it XMM-Newton\/}
observations (Dewangen et al. 2004).  This ULX is also embedded in a
spectacular nebula, visible in both optical emission lines (Pakull \&
Mirioni 2003, Kaaret, Ward \& Zezas 2004, Lehmann et al. 2005) and
radio continuum (Miller, Mushotzky \& Neff 2005).  Crucially, the flux
in the high-excitation He{\sc II} $\lambda4686$ line around the
position of the ULX implies a lower bound on its isotropic X-ray
luminosity of $4 - 6 \times 10^{39}$ erg s$^{-1}$ (Kaaret, Ward \&
Zezas 2004).  Hence Ho II X-1 is one of the strongest candidate IMBHs
in the local universe.

\section{A long look at Ho II~X-1 with {\it XMM-Newton\/}}

Ho II~X-1 was scheduled for a 112-ks {\it XMM-Newton\/} observation on
2004 April 15 (observation ID 0200470101), though the onset of intense
background flaring in the later stages of the observation caused it to
be truncated to no more than $\sim 80 - 95$ ks exposure per
instrument.  The pn was operated in full-field mode, whilst both MOS
detectors utilised the large window mode, and the medium filter was
used for all EPIC detectors.  Both the RGS and OM were in standard
modes, in the latter case monitoring Ho II with a series of UVW1
observations (though we defer discussion of this data to a later
work).

Data for the analysis were extracted from the pipeline product event
lists using the appropriate {\it XMM-Newton\/} \textsc{sas} version
6.0.0 tools. The pn data were filtered to leave only events with {\bf
flag=0} and {\bf pattern $\leq$ 4}; MOS data were filtered for {\bf
pattern $\leq$ 12} and the {\bf \#XMMEA\_EM} flag.  We quantified the
background flare contamination by examining a 10 -- 15 keV full-field
light-curve from the pn detector.  This showed the first $\sim 45$ ks
of the observation to be relatively clean, with only a few minor
flares.  However, subsequent to that strong background flaring
dominated the light-curve.  In fact, a good time intervals (GTI) file
selected from this light-curve for quiescent background times (count
rate $< 0.75$ count s$^{-1}$) leaves only 38.1 ks clean data for the
whole observation.

\subsection{Spectral analysis}

The GTI file was used to produce the cleanest possible imaging and
spectral products for the dataset.  Ho II X-1 is clearly detected in
the data, with a combined EPIC (0.3 -- 10 keV) count rate of $\sim 4
\rm ~count~s^{-1}$.  EPIC spectra and response matrices were generated
using the \textsc{especget} tool.  MOS and pn source spectra were all
extracted from an 11-pixel ($\equiv 44$ arcsec) radius circular
aperture over the position of Ho II X-1\footnote{We limit ourselves to
this relatively small aperture to avoid extracting data over pn chip
gaps, though we note that $> 87\%$ of the source flux lies within this
radius.}.  A background spectrum for the pn was extracted from a
slightly larger circular aperture on an adjacent chip, at a similar
distance from the chip readout nodes as Ho II X-1.  MOS background
data were taken from two small circular apertures at the bottom-right
and -left of each large window.  RGS source spectra, background
spectra, and response matrices were generated using the standard
\textsc{sas} pipeline script, \textsc{rgsproc}.  All spectral data
were grouped to a minimum of 20 counts per bin, in order to provide
adequate statistics for spectral fitting using $\chi^{2}$ minimization
within \textsc{xspec} version 11.3.1.

\subsubsection{RGS observations} 

\begin{figure}
\includegraphics[clip,width=60mm,angle=-90]{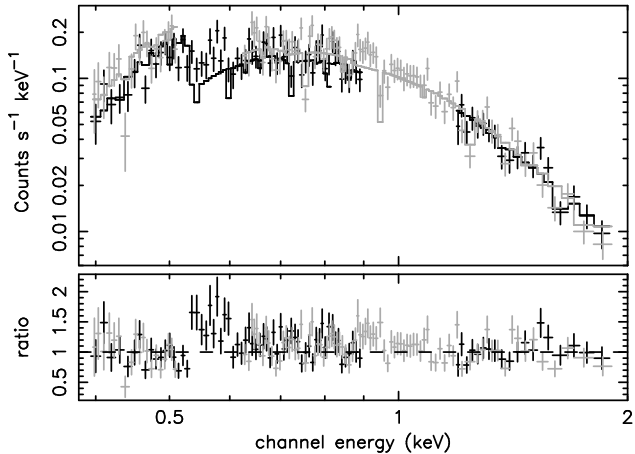}
\caption{Soft X-ray spectra of Holmberg II X-1 obtained with the
Reflection Grating Spectrometer (RGS) on-board XMM-Newton. RGS 1 data
are shown in black, RGS 2 in grey. The model fitted to the data (solid
line) is an absorbed power-law, with solar abundances, as described in
the text.  The residuals to this model (lower panel) show an excess of
counts between 0.5 and 0.6 keV.}
\label{rgs}
\end{figure}

The data extraction produced a total of 7088 source counts for both
RGS modules combined, integrated over the whole (flare-filtered)
exposure.  The resulting spectrum and residuals, compared to the
best-fit absorbed power-law model, are shown in Figure~\ref{rgs}.  We
note that this is the first reasonable quality, high resolution X-ray
spectrum of an ULX.  A fit statistic of $\chi^{2} = 430.4$ for 367
degrees of freedom (dof) is obtained for the absorbed power-law model,
where the absorption is modelled by a solar-abundance WABS model
(based on the absorption cross-sections of Morrison \& McCammon 1983),
with a rejection probability ($P_{rej}$) of 99\%.  The photon index
over the 0.3 -- 2.0 keV band was $\Gamma=2.6\pm0.2$, with an
absorption column of $N_{\rm H}=1.9\times10^{21}$~cm$^{-2}$,
significantly above the known Galactic column density towards
Holmberg~II of $N_{\rm H}=3.41\times10^{20}$~cm$^{-2}$ (Dickey \&
Lockman 1990). The measured flux over the 0.3 -- 2.0 keV RGS band is
$3.7\times10^{-12}$~ergs~cm$^{-2}$~s$^{-1}$ corresponding to an
unabsorbed luminosity of $1.2\times10^{40}$~ergs~s$^{-1}$ at 3.2 Mpc.


The residuals to the absorbed power-law fit show a clear excess of
counts between 0.5 -- 0.6\,keV, blue-wards of the neutral O edge. This
excess can be fitted with a single, broadened emission line centered
at $569\pm9$~eV, with an equivalent width of $28\pm12$~eV and a
velocity width of $\sigma=17\pm5$~eV (8900 km\,s$^{-1}$). The fit
statistic then improves to $\chi^{2} / {\rm dof} = 408.0/364$.  The
line energy is consistent with emission from a blend of lines from the
He-like O VII triplet (at 561\,eV -- 574\,eV). We therefore refitted
this excess with three narrow, unresolved lines corresponding to the
forbidden, intercombination and resonance lines from the O VII
triplet. Two lines are detected, the forbidden line measured at
$563\pm2$~eV and the resonance line at $577\pm2$~eV, both with
equivalent widths of $\sim7$~eV, $\chi^{2}=410.5$ for 363 dof. The
flux in either line is $9\pm5\times10^{-5}$~photons~cm$^{-2}$~s$^{-1}$
corresponding to a line luminosity of
$\sim1\times10^{38}$~erg~s$^{-1}$.


One plausible explanation for the origin of the line emission is from an
optically-thin thermal plasma, as has previously been suggested for this
source (Dewangan et al. 2004).  If we fit the \textsc{mekal} model in
\textsc{xspec} to the spectrum, in addition to the underlying absorbed
power-law continuum, then we can place a limit on the contribution of
such a plasma to the soft X-ray emission in HoII X-1. For solar
abundances, it is found that the 90\% confidence limit on the plasma
luminosity is $<3.6\times10^{38}$~erg~s$^{-1}$, consistent with the
value found by Dewangan et al. (2004), whilst the plasma temperature
is constrained at ${\rm kT}=175\pm25$~eV. Thus an optically thin
plasma is not likely to contribute more than 3\% of the total soft
X-ray flux in Ho II X-1.

Alternatively, the excess of counts above 0.5\,keV may arise from
incorrectly modeling the neutral O edge in the absorber model. To test
this assumption, we refitted the absorbed power-law model using the
TBVARABS absorption model, based on the improved abundances and
cross-sections tabulated by Wilms, Allen \& McCray (2000).  We find
that the O abundance is now $0.56\pm0.29$ with respect to solar, while
the column density (in excess of the Galactic value of
$3.4\times10^{20}$~cm$^{-2}$) is now $N_{\rm
H}=1.6\pm0.3\times10^{21}$~cm$^{-2}$. The fit statistic is improved
significantly, with $\chi^{2}/{\rm dof}=384.8/365$, i.e. a rejection
probability of $\sim77$\%, so formally acceptable. The decrease in
$\chi^{2}$ for this fit is due to both the lower O abundance value
relative to solar and the lower absolute abundance of O tabulated by
Wilms, Allen \& McCray (2000). The underlying photon index is then
slightly flatter ($\Gamma=2.4\pm0.2$), whilst the absorption corrected
luminosity in the 0.3 -- 2.0 keV band is
$1.0\times10^{40}$~erg~s$^{-1}$.  In this scenario no O VII emission
is formally required, as the depth of the O I edge is reduced,
resulting in little excess emission between 0.5 -- 0.6 keV.  That Ho
II X-1 lies in a low-metallicity environment should perhaps come as no
surprise; the O/H ratio for the Holmberg II galaxy is $\sim 10\%$ of
the solar abundance (Hidalgo-G{\'a}mez, S{\'a}nchez-Sacedo \& Olofsson
2003)\footnote{The somewhat higher metallicity suggested by our RGS
data in the local environment of Ho II X-1 may be a consequence of the
local star formation and/or the progenitor star for the compact object
in the ULX.}.  Indeed, Soria et al. (2005, and references therein)
suggest that low metallicity environments could favour the formation
of larger black holes ($\approx 50$ M$_{\odot}$), and hence
preferentially host ULXs.  Sub-solar abundance absorption along our
line-of-sight to Ho II X-1 therefore seems the most likely explanation
for the apparent 0.5 -- 0.6 keV excess in the RGS spectrum.

\subsubsection{EPIC observations}

\begin{table}
\flushleft
\begin{minipage}{80mm}
 \caption{{\it XMM-Newton\/} EPIC spectral fits.}
 \begin{tabular}{lccc}
\hline
Spectral	& \multicolumn{3}{c}{Absorption model} \\
parameters 	& \textsc{wabs}	& \textsc{tbvarabs}	&\textsc{tbvarabs} \\
		& (solar)	& (solar)	& ($0.56 \times$solar) \\\hline
\multicolumn{4}{c}{Model 1 : power-law continuum} \\
$N_{\rm H}~^{a}$ & $1.58 \pm 0.04$	& $1.86 \pm 0.05$	& $2.83 \pm 0.08$ \\
$\Gamma~^{b}$ 	& $2.79 \pm 0.02$	& $2.74^{+0.01}_{-0.02}$& $2.71 \pm 0.02$\\
$\chi^{2}$/dof 	& $1473.4/1221$		& $1376.2/1221$		& $1364.0/1221$ \\
$P_{rej}$	& $100\%$		& $99.88\%$		& $99.75\%$ \\
\multicolumn{4}{c}{Model 2 : accretion disc + power-law continuum} \\
$N_{\rm H}~^{a}$ & $1.66^{+0.10}_{-0.09}$& $1.77 \pm 0.09$	& $2.63^{+0.14}_{-0.15}$ \\
$kT_{in}~^{c}$ 	& $0.20 \pm 0.02$	& $0.25^{+0.02}_{-0.03}$& $0.27 \pm 0.03$ \\
$A_{DBB}~^{d}$ 	& $110^{+76}_{-42}$	& $23.3^{+15.4}_{-8.5}$	& $11.5^{+7.3}_{-4.4}$ \\
$\Gamma~^{b}$ 	& $2.64 \pm 0.03$	& $2.62^{+0.03}_{-0.04}$& $2.60 \pm 0.04$ \\
$\chi^{2}$/dof 	& $1315.1/1219$		& $1324.1/1219$		& $1335.0/1219$	\\
$P_{rej}$	& $97.2\%$		& $98.1\%$		& $98.9\%$ \\
$F_{disc}~^{e}$	& $0.09$		& $0.07$		& $0.06$ \\
\multicolumn{4}{c}{Model 3 : accretion disc + Comptonised corona} \\
$N_{\rm H}~^{a}$ &$0.65^{+0.13}_{-0.15}$& $0.59^{+0.11}_{-0.13}$& $0.82^{+0.16}_{-0.18}$ \\
$T_{max}~^{f}$ 	& $0.18^{+0.02}_{-0.01}$& $0.19^{+0.02}_{-0.01}$& $0.19 \pm 0.01$ \\
$kT_e~^{g}$ 	& $2.9^{+1.9}_{-0.7}$	& $2.8^{+1.5}_{-0.6}$	& $2.7^{+1.5}_{-0.5}$ \\
$\tau~^{h}$ 	& $4.1^{+0.7}_{-0.8}$	& $4.3^{+0.8}_{-1.1}$	& $4.4^{+0.8}_{-1.2}$ \\
$\chi^{2}$/dof 	& $1286.4/1218$		& $1289.9/1218$		& $1291.7/1218$	\\
$P_{rej}$	& $91.5\%$		& $92.6\%$		& $93.0\%$ \\
$F_{disc}~^{e}$	& $0.13$		& $0.15$		& $0.15$ \\
\hline
\end{tabular}
{\footnotesize The models used (subject to the tabulated absorption
models) were, in \textsc{xspec} syntax: (1) PO; (2) DISKBB + PO; (3)
DISKPN + COMPTT.  We quote errors at the 90\% level for one
interesting parameter.  The parameters are: $^{a}$ Absorption column
external to our Galaxy in units of $10^{21}$ atom~cm$^{-2}$. $^{b}$
Power-law continuum photon index.  $^{c}$ Inner-disc temperature
(keV). $^{d}$ Disc model normalisation.  $^{e}$ Fraction of observed
0.3 -- 10 keV flux in the accretion disc component.  $^{f}$ Maximum
temperature of the disc (keV). $^{g}$ Coronal temperature
(keV). $^{h}$ Coronal optical depth. }
\end{minipage}
\label{specfits}
\end{table}

EPIC spectral data were fit over the 0.3 -- 10 keV band, though within
this we only fit the MOS over 0.3 -- 6 keV (above this the data were
noise-dominated), and the pn over 0.7 -- 10 keV (known calibration
inaccuracies between 0.5 -- 0.7 keV are particularly prominent in this
spectrally-soft dataset).  We allowed for residual calibration
uncertainties between the three detectors in our fits by introducing a
constant term into our spectral fitting; in practice the variation
between the detectors was $< 6\%$.  Each model we fit to the data was
subject to a fixed Galactic absorption column of $3.41 \times
10^{20}$~cm$^{-2}$, modelled using a solar-abundance \textsc{xspec}
WABS cold absorption model.

We began, as with the RGS analysis, by fitting an absorbed power-law
continuum model, using a second WABS component to fit additional
line-of-sight absorption to Ho II X-1 (which is presumably intrinsic
to either the ULX itself, or its host regions in Holmberg II).  This
fit was rejected at high significance (see Table 1), though its best
fit parameterisation was similar to that of the narrower-band RGS
result ($\Gamma \sim 2.8$, $N_{\rm H} \sim
1.6\times10^{21}$~cm$^{-2}$).  However, we also attempted a second
fit, this time using the TBVARABS model, with the actual abundance of
the absorbing medium set to 0.56 times solar as suggested by the RGS O
I edge feature.  This produced a vastly improved fit ($\Delta\chi^2 =
109$ for the same dof).  This improvement was predominantly down to
the use of the TBVARABS model itself, which includes (amongst other
improvements) a much lower absolute O abundance than WABS.  We
tabulate the results of these two fits, plus a TBVARABS fit set to
solar abundance, in Table 1.  However, even though the fit is greatly
improved by the use of TBVARABS, a simple power-law continuum fit is
still significantly rejected.

We next attempted two-component models consisting of a power-law
continuum plus an accretion disc model, similar to previous work on
ULXs, and the empirical model for X-ray emission from a black hole
X-ray binary.  As with previous work, we utilised the \textsc{xspec}
DISKBB model, based on the multi-colour disc blackbody approximation
(e.g. Mitsuda et al. 1984).  This model provides an improvement to the
fit; however, as Table 1 shows the magnitude of this improvement is
highly dependent upon the choice of absorption model, with the solar
abundance WABS fits showing the largest improvement ($\Delta\chi^2 =
158$ for 2 extra dof), and the best actual fit, whilst the most
physical absorption model (sub-solar TBVARABS) shows the smallest
improvement ($\Delta\chi^2 = 29$ for 2 extra dof) and worst fit.  In
all cases the DISKBB temperature was relatively cool ($\sim 0.2 -
0.27$ keV), albeit hotter than found in the previous observations of
Ho II X-1 by Dewangen et al. (2004), and the power-law continuum
remains steep ($\Gamma \sim 2.6$)\footnote{We note that we also
attempted the ``non-standard'' form of the DISKBB + power-law model,
as described by Stobbart et al. (2004) and Roberts et al. (2005).
This model was only marginally less successful than the standard
usage, by $\Delta\chi^2 \sim 15$ for TBVARABS models, with best fits
showing $kT_{in} \sim 2$ keV for the accretion disc, and $\Gamma \sim
3$ for the power-law continuum.}.

\begin{figure}
\includegraphics[clip,width=55mm,angle=-90]{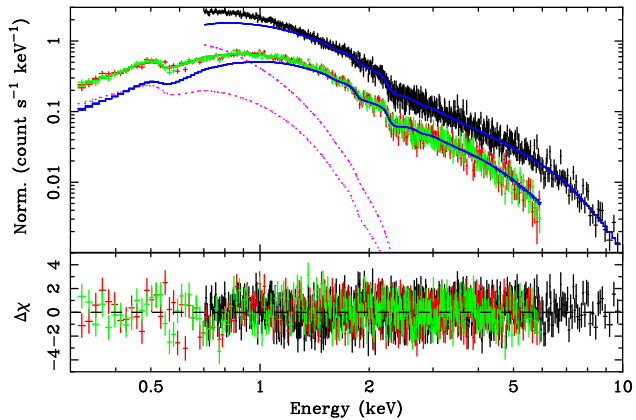}
\caption{EPIC spectral data for Ho II X-1, plotted with the best-fit
spectral model (3, with sub-solar abundance) and
$\Delta\chi$-residuals.  The pn data are shown in black, with MOS1 and
MOS2 in grey.  The contributions of the DISKPN and COMPTT components
are shown by the dotted and solid lines respectively.}
\label{epicspec}
\end{figure}

The DISKBB + power-law model is only a first-order approximation of
the physics of an accretion disc plus corona system.  Hence we next
tried a more physically-realistic, self-consistent model, namely the
DISKPN + COMPTT combination (see Gierli{\'n}ski,
Machiolek-Nied{\'z}wicki \& Ebisawa 2001; Titarchuk 1994 for
descriptions of these models) that describes an accretion disc
modified for e.g. relativistic effects, and a Comptonised corona.  We
link the two models by setting the seed photon temperature of the
corona, $T_0$, to the maximum temperature of the disc $T_{max}$, and
we set the inner radius of the disc to the innermost stable circular
orbit ($= 6 R_g$).  This model provides a statistically-acceptable
solution to the spectrum, regardless of the absorption model used (as
the modelled absorption is relatively low, there is little difference
between the models in this case).  Whilst the disc again appears cool
($T_{max} \sim 0.2$ keV), the best-fit model shows a remarkable
corona, in that it appears both quite cool ($kT \sim 2.7$ keV) and
optically thick ($\tau \sim 4.4$).  We show the best fit spectrum,
with the contributions from the two components decomposed, in
Fig.~\ref{epicspec}, and tabulate the interesting parameters in Table
1.  These models provide an observed 0.3 -- 10 keV flux measurement of
$\sim 6.7 \times 10^{-12}$ erg cm$^{-2}$ s$^{-1}$, which converts to
an intrinsic luminosity of $1.0 \times 10^{40}$ erg s$^{-1}$ for a
distance of 3.2~Mpc (this intrinsic luminosity is similar to the RGS
estimate - taken from a much narrower band - as the absorption
correction is considerably smaller in our more complex EPIC
modelling).

We have also investigated whether an optically-thin corona also
provides a reasonable solution to the spectral data, as one might
expect given the DISKBB + power-law fits to the data.  This was done
by attempting two new fits with the coronal temperature set to 50 and
100 keV respectively, indicative of the range seen in Galactic black
hole binaries, in the DISKPN + COMPTT models.  These fits did indeed
provide reasonable solutions to the data, with a best-fitting $\tau <
0.25$ (the other parameters did not change substantially from the
values in Table 3 for the optically-thick solution) and $P_{rej} \sim
95\%$.  However, they were still worse than the optically-thick
solutions by $\Delta\chi^2 \sim 12 - 16$ for one extra dof, which
equates to a significance of $3 - 4\sigma$ according to the
F-statistic.  Therefore we regard the presence of an optically-thick
corona as far more likely than an optically-thin one, though we
obviously cannot exclude the latter.

Finally, as Fig.~\ref{epicspec} demonstrates, the data show no obvious
signs of emission lines including Fe K.  We place a limit on
the possible contribution of a narrow, neutral (i.e. 6.4 keV) Fe K
line by adding a Gaussian line model to the best-fit spectra; from
this we derive a 90\% upper-limit of $25$ eV on the equivalent width
of such a feature.

\subsection{Temporal analysis}

\begin{figure}
\includegraphics[clip,width=44mm,angle=-90]{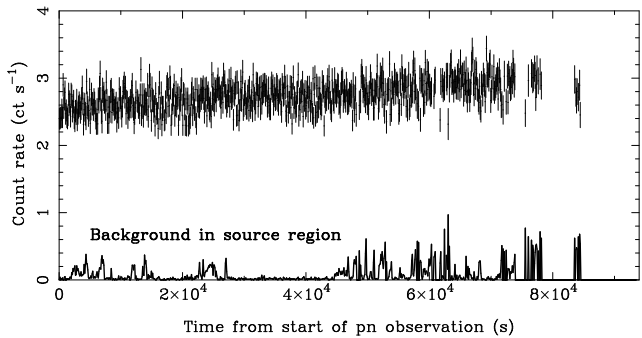}
\caption{EPIC pn 0.3 -- 6 keV light-curve for the long-look at Ho II
X-1.  The data is binned to 100-s intervals, and background
subtracted.  We also show the relative intensity of the background
emission in the same band, for the same aperture.  The gaps in the
data are caused by the telemetry buffers filling up due to the high
background flaring rate.  For clarity we only show data bins where the
we have 75-s or more of exposure.}
\label{pnlc}
\end{figure}

Temporal data for Ho II X-1 were extracted from the EPIC detectors in
the same circular apertures as used in the spectral analysis.
Background data were extracted from same-sized apertures on the same
CCD chip as the source.  Given the relative spectral softness of the
ULX, and the comparatively hard X-ray nature of background flaring
detected by {\it XMM-Newton\/}, we maximised our temporal
signal-to-noise by selecting only data in the 0.3 -- 6 keV range.  The
use of relatively small apertures, in combination with excluding hard
X-ray data, ensured that the source signal per unit area dominated
above even the strongest background flaring.  We demonstrate this for
the pn data in Fig.~\ref{pnlc}, which shows the full light-curve for
this observation of Ho II X-1 binned to 100~s resolution.

Although Ho~II~X-1 shows substantial variability on time-scales of
months-years (e.g. Dewangan et al. 2004), the XMM-Newton light curve
shows very little variability on time-scales of less than a day
(barring a very weak monotonic trend throughout the observation).  To
quantify and constrain the variability more precisely, we extracted a
0.3--6.0~keV EPIC-pn light curve at the highest possible resolution
(73.4~ms for full-frame mode), corresponding to the first 48.2~ks of
data, which is free of telemetry dropouts and significant background
flaring.  We split the light curve into five equal segments and
measured the PSD of each segment (with power determined in units of
$(rms/mean)^{2}$), then merged and averaged the resulting PSDs by
binning in frequency, so that each bin corresponds to a factor of 1.1
increase in frequency (with the proviso that each bin contains at
least 20 individual PSD measurements).  The errors on the averaged
powers were calculated from the rms scatter on individual PSD
measurements in each frequency bin.  The resulting PSD was found to be
completely flat, consistent with variations being due to photon
counting statistics only.  To constrain the strength of any intrinsic
source variability, we used {\sc xspec} to fit the PSD.  We obtain an
excellent fit ($\chi^{2}=74.3$ for 88 degrees of freedom) for a
constant power of $0.793\pm0.002$~Hz$^{-1}$ (errors are 90\%
confidence).  The expected power level due to Poisson noise for the
observed count rate is 0.81~Hz$^{-1}$.  The fact that the observed
constant power is {\it smaller} than expected may be related to weak
instrumental effects (eg. pileup, deadtime).

\begin{figure}
\begin{center}
\includegraphics[clip,width=50mm,angle=270]{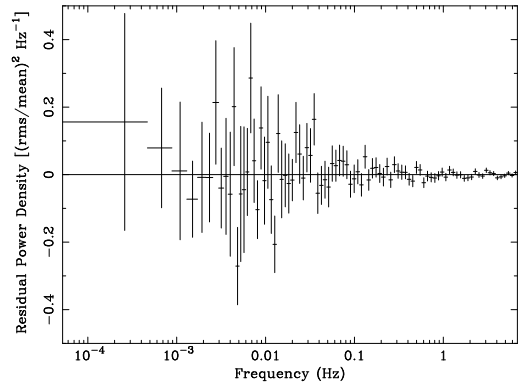}
\end{center}
\caption{Residuals from the observed PSD fitted with a constant.}
\label{psdplot}
\end{figure}

The fit residuals relative to the best-fitting constant are plotted in
Figure~\ref{psdplot}.  No systematic features are apparent.  At first
glance, the data might suggest that the fastest intrinsic variability
of the source occurs on time-scales longer than observed ($>$day),
thus supporting the idea that the source is an intermediate mass black
hole.  However, AGN with black hole masses $\sim10^{6}$~M$_{\odot}$
show significant variability on time-scales of minutes-hours
(e.g. M$^{\rm c}$Hardy et al. 2004), and we would therefore expect the
same if the source is an intermediate mass black hole, unless it
occupies a state with unusually low variability power.  To demonstrate
this fact more quantitatively, we attempted to fit the PSD with a
constant plus a broken power-law of low-frequency slope $-1$, and
slope $-2$ above the break, similar to the PSD shape observed in
several AGN and Cyg~X-1 in the high/soft state (e.g. M$^{\rm c}$Hardy
et al. 2004, 2005; Uttley \& M$^{\rm c}$Hardy 2005).  In units of {\it
power}$\times${\it frequency} ($\nu P(\nu)$), a PSD slope of $-1$
corresponds to a constant value $C_{1/f}$ (which when multiplied by
natural log of 10 gives the power per decade in frequency).  In the
AGN observed to date, and in Cyg~X-1 and other black hole XRBs, this
constant value is $\sim 0.005-0.03$.

For a break frequency of $10^{-4}$~Hz, which is even lower than those
observed in several AGN (and corresponds to a black hole mass of
$\sim10^{6}$~M$_{\odot}$, assuming linear scaling of characteristic
time-scales from the $\sim10$~Hz break seen in Cyg~X-1), the
99~per~cent upper limit to $C_{1/f}$ is $2.8\times 10^{-4}$.  For a
PSD break at 1~Hz or higher, which would correspond to black hole
masses of 100~M$_{\odot}$ or less, the limit is
$C_{1/f}<1.4\times10^{-4}$, which corresponds to an intrinsic
fractional rms over the entire observed PSD of less than 1 per cent.
In some black hole XRBs in the high/soft state, the amplitude of
variability is diluted significantly by the presence of strong
constant blackbody emission from the disc which dominates the X-ray
spectrum and dilutes the variability in the power-law component.
However, our spectral fits show that the X-ray emission we observe is
{\bf not} dominated by the putative disc component (that accounts for
only $\sim 20\%$ of the observed counts in the 0.3 -- 6 keV pn data),
hence we conclude that the source cannot be accreting in a state with
timing properties similar to the high/soft state of Cyg~X-1 or AGN
with good PSD measurements.

We next consider the possibility that the PSD is `band-limited'
(e.g. see van der Klis 2005), that is, the source does vary
significantly but that variability is confined to a limited frequency
range (perhaps a decade or two, or even smaller if the variability is
dominated by strong QPOs).  Such band-limited PSDs are observed in
black hole XRBs in the low/hard state (where relatively hard
($\Gamma<2$) power-law emission dominates the X-ray spectrum) and also
the very-high state, where the spectrum shows strong, steep power-law
emission, similar to that observed here (see McClintock \& Remillard
2003 for more detailed descriptions of these states).  The
high-frequency cut-offs in these PSDs typically occur around 1--10~Hz,
and simple mass-scaling arguments then imply that the peaks in
band-limited power should be sampled by our observed frequency range,
for intermediate mass or stellar mass black holes.  In that case, we
may not detect these peaks due to lack of sensitivity, which decreases
at high-frequencies due to the Poisson noise contribution to the
PSD\footnote{Note that for a band-limited component with constant rms
and constant width in decades, the peak power also remains constant in
$\nu P(\nu)$ units, but the noise level increases linearly with
frequency, hence the reduction in sensitivity at high frequencies.}.

\begin{figure*}
\begin{center}
\includegraphics[clip,width=50mm,angle=270]{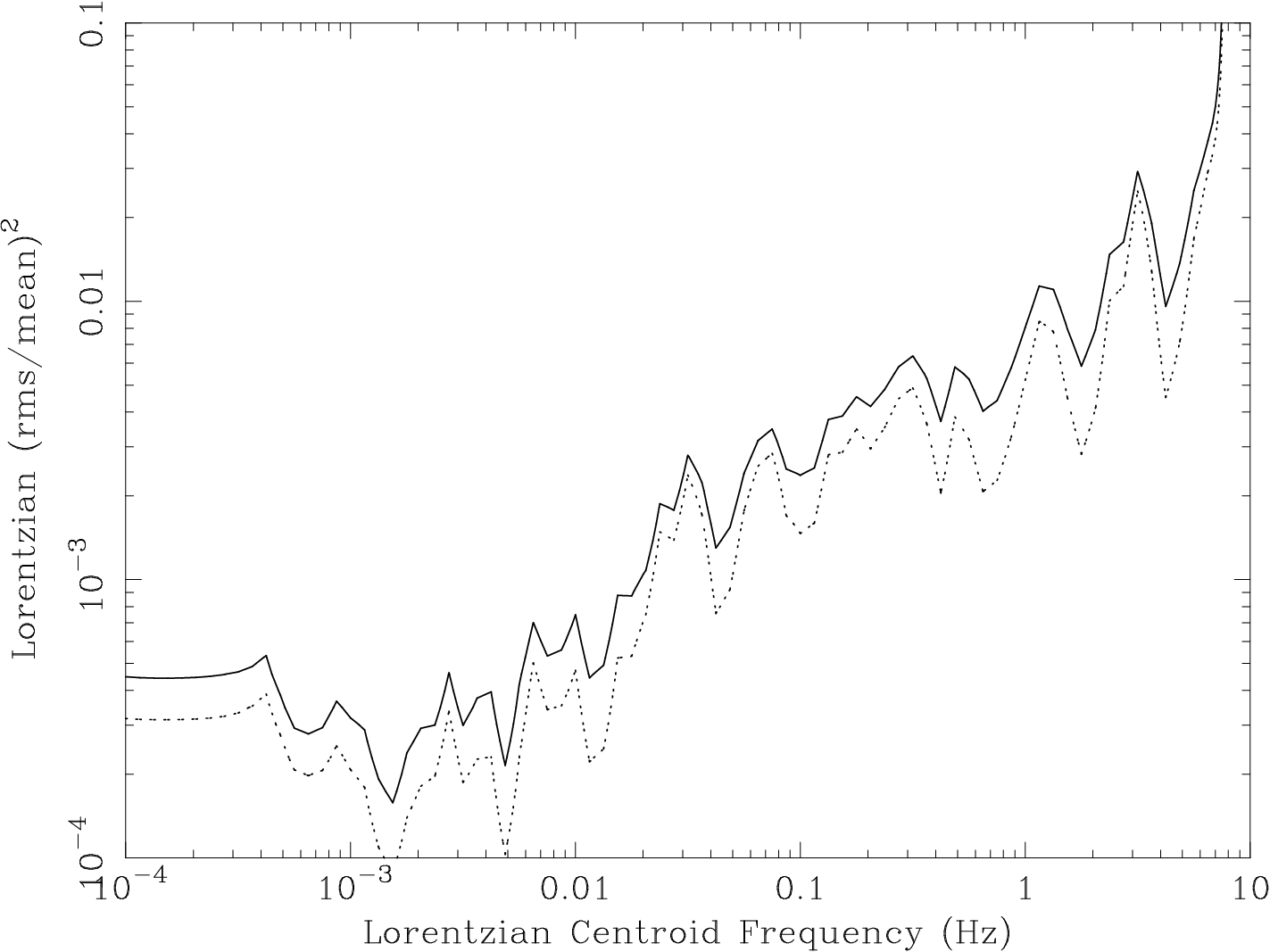}\hspace*{1cm}
\includegraphics[clip,width=50mm,angle=270]{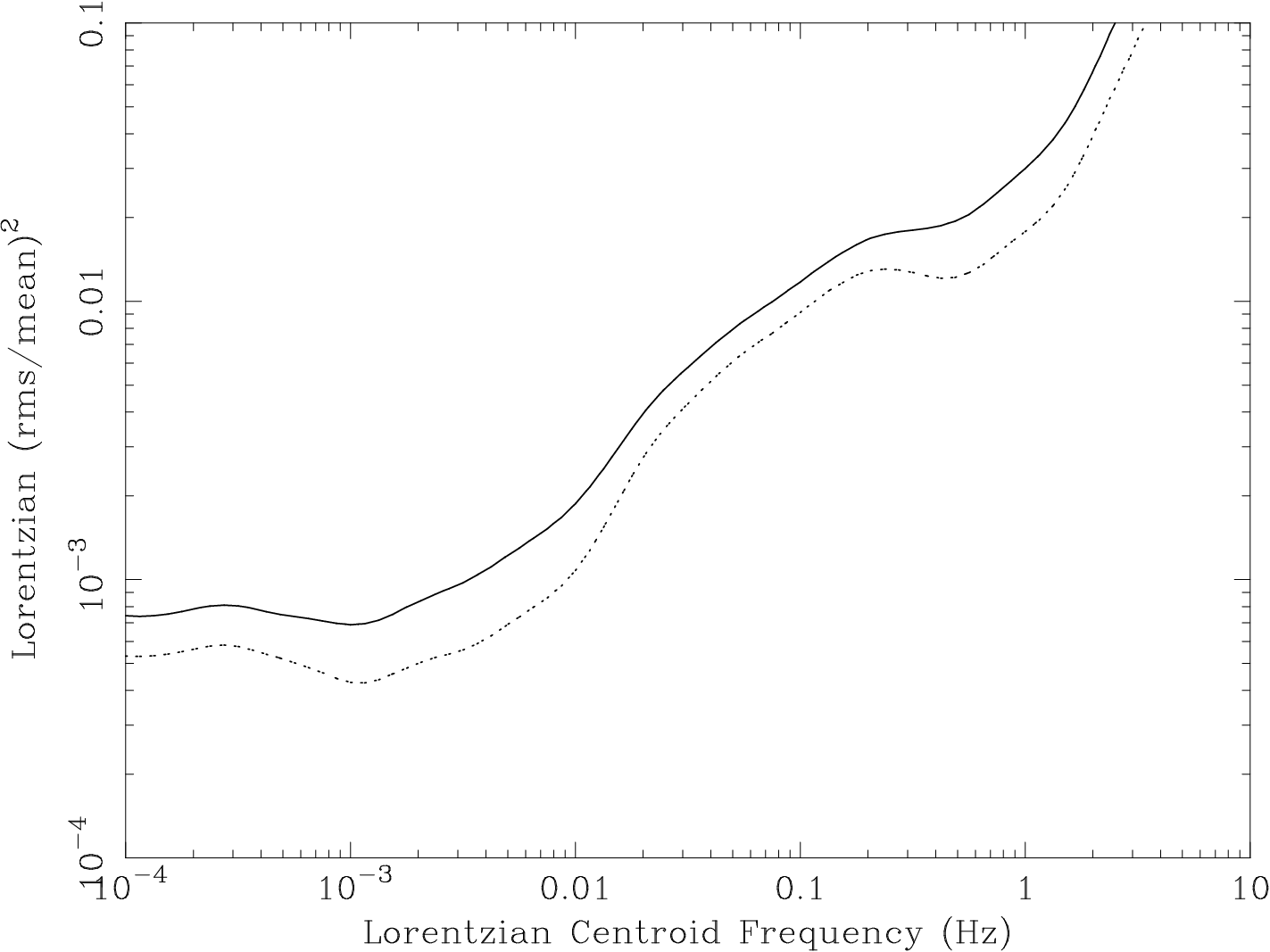}
\end{center}
\caption{Contour plots showing upper limits
on the amplitudes of broad ($Q=0.5$, left) and narrow ($Q=5$,
right) Lorentzian components in the observed PSD (see text for details)}
\label{lorcontours}
\end{figure*}

To determine upper limits on the amplitude of band-limited noise which
may be hidden in the PSD, we fitted the PSD with a constant plus
Lorentzian model.  The Lorentzian represents the band-limited
contribution: it's coherence or $Q$-value (the centroid frequency
divided by the full width at half-maximum) can be changed to represent
different observed types of band-limited noise or QPOs.  For the
incoherent band-limited noise typically observed in black hole XRBs in
either state, we assumed $Q=0.5$ (e.g. see Pottschmidt et al. 2003).
For the strong, sharp QPOs which are also often observed in very-high
states (and which would be easier to detect than band-limited noise of
the same fractional rms), we assumed $Q=5$ (e.g. see Remillard et
al. 2002, van der Klis 2005).  For these two types of Lorentzian, we
fitted the PSD to obtain the 90 and 99~per~cent upper limits on
variability amplitude (in terms of the square of fractional rms
contributed by the Lorentzian), which is permitted as a function of
Lorentzian centroid frequency.  The resulting contour plots are shown
in Fig.~\ref{lorcontours}(a,b).  The limits we can place on the
presence of significant variability from band-limited noise are very
stringent.  For example, the typical observed fractional rms in a
single Lorentzian is $\sim10$ per cent, i.e. fractional rms-squared
$\sim0.01$, which is ruled out at 99 per cent confidence for
frequencies less than $\sim0.05$~Hz and $<1$~Hz for $Q=0.5$ and $Q=5$
respectively.

Assuming linear scaling of time-scales with black hole mass, these
limits on the allowed frequencies of the expected band-limited noise
suggest that the black hole mass should be of order $100$~M$_{\odot}$
or less, assuming that X-ray timing behaviour is similar for all black
holes, regardless of black hole mass.  In fact the lack of observed
variability is consistent with the PSD being similar to that observed
in certain states of the microquasar GRS~1915+105, which, although
famous for showing highly variable and complex light curves, also
shows states (the $\chi$ class of behaviours, see Belloni et al. 2000)
where the long-term X-ray variability (on time-scales up to months)
can be remarkably constant, with all the variability power being
concentrated at high frequencies ($\sim 1$~Hz or higher).



\section{Discussion}

Our analysis of the RGS data for Ho II X-1 shows that it is likely
that the X-ray emission we observe from the ULX is subject to
absorption by a sub-solar abundance medium, almost certainly within
the Holmberg II galaxy, and possibly in the immediate environment of
the ULX itself.  Interestingly, this result has implications for
parameterisation of the EPIC X-ray spectrum.  In particular, the
mis-modelling of the neutral O edge as an solar-abundance absorber
leads to an over-estimation of the magnitude of the soft excess in the
empirical models (power-law continuum and power-law + DISKBB models)
of the data.  The main implication of this is that it can lead to an
over-estimate of the mass of the putative IMBH derived from the DISKBB
parameters.  Using equations (5) and (8) of Makishima et al. (2000),
and the best-fit parameters from the empirical power-law + DISKBB
fits, the IMBH mass is estimated as $624$
$[cos(i)]^{-{{1}\over{2}}}\alpha^{-1}$ \mdot for the solar-abundance
WABS absorption model\footnote{$i$ is the inclination of the accretion
disc to the line-of-sight, and $\alpha$ relates to the spin of the
black hole ($\alpha = 1$ in the Schwarzschild metric, and $\alpha < 1$
for Kerr black holes).}.  When the sub-solar TBVARABS model is used
instead, the black hole mass reduces to $206$
$[cos(i)]^{-{{1}\over{2}}}\alpha^{-1}$ \mdot, down to $\sim 33\%$ of
the previous estimate.  This explicitly demonstrates the sensitivity
of the estimated IMBH masses to the assumed absorption model, and may
imply that previous masses in the literature derived by this technique
are somewhat over-estimated.

Furthermore, we caution that mass estimates derived from modelling the
soft-excess are likely very highly uncertain.  Gierlinski and Done
(2004) in a systematic study of all radio-quiet PG quasars observed
with {\it XMM-Newton\/}, found that the temperature of the soft excess
was remarkably constant despite the 3 orders of magnitude range in
black hole mass (their Figure 1, panel b) and order of magnitude range
in $L/L_{\rm edd}$. This apparent insensitivity of the soft-excess to
either mass or luminosity has been noted before (Walter and Fink 1993;
Czerny et~al 2003; Porquet et~al. 2004), and casts strong doubt on the
interpretation of soft excesses as an accretion disc signature in
QSOs.

If the same uncertainty is applied to ULXs, we may require a different
explanation for the origin of the soft excess.  One possibility is an
optically thick, low temperature photosphere.  This phenomenon should
naturally arise in sources accreting at or close to the Eddington
limit (King \& Pounds 2003) without appealing to large black hole
masses.  In the King and Pounds model, any black hole accreting at
close to Eddington will produce a strong Compton-thick outflow with an
effective photospheric size of a few tens of $R_{\rm Schw}$ from the
central source.  Using their equation 18, and scaling down their
numbers for the PG quasar 1211+143, then for a 10$M_{\odot}$ black
hole accreting at close to Eddington we predict an outflow rate of
$\sim 10^{-7}M_{\odot}$/yr, which implies an effective blackbody
temperature for the photosphere of a few $\times 100$~eV, similar to
the value we measure for Ho II~X-1.  However, we note that the largest
challenge for this model is to explain how we still observe the hard
X-ray component through the Compton-thick outflow (cf. Miller, Fabian
\& Miller 2004).

If we return to the standard assumption - namely, the soft excess does
originate in the thermal emission of an accretion disc - then the mass
estimates assume that the disc is in the ``high'' spectral state
(i.e. radiating from the innermost stable circular orbit around the
black hole, $\equiv 6R_g$ for a Schwarzschild black hole).  We might
then expect the rest of the X-ray spectrum and its variability
characteristics to conform to this state.  Unlike some other ULXs, Ho
II X-1 does display the steep power-law continuum (in the empirical
power-law + DISKBB fit), $\Gamma > 2.4$, characteristic of the high
state (cf. Roberts et al. 2005).  However, it does not display the
correct variability characteristics.  This is a particular problem as
if an IMBH is present, even with the reduced mass derived above, the
X-ray luminosity implies it is apparently accreting at $\sim 0.1 -
0.5$ L$_{Edd}$.  This accretion rate regime is observed in both AGN
and XRBs, all of which show a substantial level of variability in the
high state.  Assuming that accretion physics works similarly for
IMBHs, this apparent failure of XRB properties to ``scale-up'' to an
IMBH is a severe problem.

There is a second way in which this ULX appears ``abnormal'' compared
to Galactic XRBs in the high state.  Physical modelling of the source
spectrum reveals its putative coronal component is probably both
optically-thick and cool ($\tau \sim 4, kT \sim 3$ keV), in stark
contrast to the optically-thin, hot ($\tau \sim 1, kT \sim 50 - 100$
keV) coronae found around Galactic black holes.  The coronal X-ray
emission also quantitatively dominates the observed X-ray emission,
contributing $> 80\%$ of the 0.3 -- 10 keV flux, unlike Galactic black
holes in the high state (cf. M${\rm ^c}$Clintock \& Remillard
2005\footnote{A caveat must be that if an IMBH is present, then to
provide a fair comparison one would have to use a softer X-ray band
than the 2 -- 20 keV band these authors use to define their black hole
emission states.  One might argue that our 0.3 -- 10 keV data provides
this, and is still very corona-dominated.})  In fact, an X-ray
spectrum dominated by coronal emission is more indicative of the very
high state.  Interestingly enough, an optically-thick medium could
provide a plausible means to explain the lack of any accretion disc
features (in particular an Fe K line) in the X-ray spectrum.  However,
these arguments again imply that this ULX does not appear to be
operating as a scaled-up high state XRB.  Hence models assuming a
$\sim 1000$ \mdot IMBH in the high state underlies Ho II X-1 face
considerable new challenges on the basis of our data.


Is there an alternative explanation for the spectral and temporal
characteristics of this source?  The Galactic microquasar GRS 1915+105
might hold the key.  In particular, Zhang et al. (2000) have
demonstrated that its X-ray spectrum can be modelled by a three-layer
atmospheric structure, broadly similar to the solar atmosphere.  These
three layers are: (i) a cold and optically-thick disc ($kT \sim 0.2 -
0.5$ keV); (ii) a ``warm layer'' overlaying the cold disc with $kT
\sim 1 - 1.5$ keV and $\tau \sim 10$; and (iii) a hot, optically-thin
corona ($kT \sim 100$ keV, $\tau \sim 1$), though this last component
is not always present\footnote{Broad-band spectral studies of GRS
1915+105 can require models far more complex than this simple model,
see for example Done, Wardzinski \& Gierlinski (2004); Zdziarski et
al. (2005).}.  This model has striking similarities to our physical
model, in particular a cool disc seeding a warm, optically-thick
scattering medium.  Our lack of variability could then be explained by
the non-detection (within our limited spectral range) of the
optically-thin coronal component and/or the source being in the
$\chi$-class of GRS 1915+105 behaviour.  Indeed, as the $\chi$-class
of GRS 1915+105 appears typical of the very high state (Zdziarski et
al. 2005), as is a corona-dominated X-ray spectrum, then the mass
limit derived from band-limited variability considerations may be
appropriate.  Crucially, when combined with the non-dependence of the
black hole mass on the disc temperature in this model, this allows us
a far smaller black hole than conventionally assumed for this ULX ($<
100$ \mdot).  This in turn implies that Ho II X-1 must be accreting at
(or perhaps above) the Eddington limit.  We note that theoretical
models for ULXs as stellar-mass black holes accreting from high mass
donor stars certainly permit sufficient mass transfer to fuel such
high luminosities (Rappaport, Podsiadlowski \& Pfahl 2005), and
speculate that this excessive mass transfer may also result in enough
material transfering to the corona to make it optically-thick.

Therefore we conclude that it remains plausible that Ho II X-1 may
possess a black hole no larger than a few tens of solar masses
radiating at or above the Eddington limit.

\section{Summary}

We present an unprecedented deep observation of one of the best IMBH
candidate ULXs, Ho II X-1.  This data provides a challenge to previous
results suggesting the presence of a $\sim 1000$ \mdot~ IMBH, most
notably in the lack of significant variability over the timescales
probed by our data, and the possible detection of an optically-thick
corona, both of which are inconsistent with the simple argument for
IMBHs based on the ``scaling-up'' of a Galactic XRB.  A plausible
alternative is that Ho II X-1 is behaving like GRS 1915+105 in its
$\chi$-class very high state; in this case the lack of observed
variability places a limit on the mass of the compact accretor in Ho
II X-1 at $< 100$ \mdot.

\section*{Acknowledgments}

We would like to thank the anonymous referee for a careful reading of the
manuscript and for a number of helpful suggestions which led to an improvement
in the work presented here.

M.R. Goad and T.P. Roberts acknowledge financial support from PPARC. P. Uttley
is supported by an NRC Research Associateship.  This work is based on
observations with {\it XMM-Newton\/}, an ESA science mission with instruments
and contributions directly funded by ESA member states and the USA (NASA).





\label{lastpage}


\begin{thebibliography}{}

\bibitem[\protect\citeauthoryear{}{}]{} 
Begelman M.C., 2002, ApJ, 568, L97

\bibitem[\protect\citeauthoryear{}{}]{} 
Belloni T., Klein-Wolt M., M\'{e}ndez M., van der Klis M., van
Paradijs J., 2000, A\&A, 355, 271

\bibitem[\protect\citeauthoryear{CP}{2002}]{cp1}
Colbert E.J.M., Ptak, A.F., 2002, ApJS, 143, 25	
									     
									     
\bibitem[\protect\citeauthoryear{}{}]{} 
Cropper M., Soria R., Mushotzky R.F., Wu K., Markwardt C.B., Pakull
M., 2004, MNRAS, 349, 39

\bibitem[\protect\citeauthoryear{}{}]{}
Czerny, B., Nikolajuk, M. R{\'o}za{\'n}ska, A. Dumont, A.-M. 2003, A\&A, 412,
317.

\bibitem[\protect\citeauthoryear{DMGL}{2004}]{dmgl1}
Dewangan G., Miyaji T., Griffiths R.E., Lehmann I., 2004, ApJ, 608, L57 

\bibitem[\protect\citeauthoryear{DL}{1990}]{dl1}
Dickey J.M., Lockman F.J., 1990, ARA\&A, 28, 215

\bibitem[\protect\citeauthoryear{}{}]{} 
Done C., Wardzinski G., Gierlinski M., 2004, MNRAS, 349, 393

\bibitem[\protect\citeauthoryear{}{}]{} 
Ebisawa K., Zycki P., Kubota A., Mizuno T., Wataria K., 2003, ApJ,
597, 780

\bibitem[\protect\citeauthoryear{}{}]{} 
Ebisuzaki T., et al., 2001, ApJ, 562, L19

\bibitem[\protect\citeauthoryear{EN}{1999}]{en1}
Edelson R., Nandra, K., 1999, ApJ, 514, 682.

\bibitem[\protect\citeauthoryear{}{}]{}
Fabbiano G., Trinchieri G., 1987, ApJ, 315, 46.

\bibitem[\protect\citeauthoryear{FPMD}{2000}]{fpmd1}
Ferrarese L., Pogge R.W., Peterson B.M., Merritt D., Wandel A., Joseph
C.L., 2001, ApJ, 555, L79

\bibitem[\protect\citeauthoryear{}{}]{}
Gebhardt K.,~et al., 2000, ApJL, 543, L5

\bibitem[\protect\citeauthoryear{}{}]{} 
Gierlinski M., Maciolek-Niedzwiecki A., Ebisawa K., 2001, MNRAS, 325, 1253

\bibitem[\protect\citeauthoryear{}{}]{}
Gierli{\'n}ski, M. and Done, C. 2004, MNRAS 349, L7.

		     
     
\bibitem[\protect\citeauthoryear{}{}]{} 
Hidalgo-Gamez A.M., Sanchez-Salcedo F.J., Olofsson K., 2003, A\&A, 399, 63


\bibitem[\protect\citeauthoryear{}{}]{} 
Kaaret P., Ward M.J., Zezas A., 2004, MNRAS, 351, L83

\bibitem[\protect\citeauthoryear{}{}]{}
King A.R., Davies M.B., Ward M.J., Fabbiano G., Elvis, M., 2001, ApJL,
552, L109

\bibitem[\protect\citeauthoryear{}{}]{}
King, A.R. and Pounds, K.A. 2003, MNRAS 345, 657.
									     
\bibitem[\protect\citeauthoryear{}{}]{} 
K{\"o}rding E., Falcke H., Markoff S., 2002, A\&A, 382, L13

\bibitem[\protect\citeauthoryear{}{}]{} 
Lehmann I., et al., 2005, A\&A, 431, 847


\bibitem[\protect\citeauthoryear{}{}]{}
Madau P., Rees M.J., 2001, ApJL, 551, L27

\bibitem[\protect\citeauthoryear{}{}]{} 
Makishima K. et al., 2000, ApJ, 535, 632

\bibitem[\protect\citeauthoryear{}{}]{}
Markowitz A., et al., 2003, ApJ, 593, 96
									     
\bibitem[\protect\citeauthoryear{}{}]{} 
M$^{\rm c}$Clintock J. E., Remillard R. A., 2005, in Compact Stellar
X-ray Sources, ed. Lewin W. H. G., van der Klis M., Cambridge
University Press (Cambridge), in press (astro-ph/0306213)

\bibitem[\protect\citeauthoryear{}{}]{} 
M$^{\rm c}$Hardy I. M., Papadakis I. E., Uttley P., Page M., Mason K.,
2004, 348, 783

\bibitem[\protect\citeauthoryear{}{}]{} 
M$^{\rm c}$Hardy I. M., Gunn K. F., Uttley P., Goad M. R., 2005,
MNRAS, 359, 1469

   
\bibitem[\protect\citeauthoryear{}{}]{}
Miller J.M., Fabbiano G., Miller M.C., Fabian A.C., 2003, ApJL, 585, L37

\bibitem[\protect\citeauthoryear{}{}]{} 
Miller J.M., Fabian A.C., Miller M.C., 2004, ApJ, 607, 931

\bibitem[\protect\citeauthoryear{}{}]{} 
Miller N.A., Mushotzky R.F., Neff S.G., 2005, ApJ, 623, L109


\bibitem[\protect\citeauthoryear{}{}]{} 
Mitsuda K., et al., 1984, PASJ, 36, 741

\bibitem[\protect\citeauthoryear{}{}]{}
Miyaji T., Lehmann I., Hasinger G., 2001, AJ, 121, 3041		      
									      
\bibitem[\protect\citeauthoryear{}{}]{} 
Morrison R., McCammon D., 1983, ApJ, 270, 119

									      
\bibitem[\protect\citeauthoryear{}{}]{}
Pakull M.W., Mirioni, L., 2002, preprint ({\tt astro-ph/0202488})
									      

\bibitem[\protect\citeauthoryear{}{}]{}
Pourquet, D., Reeves, J.N., O'Brien, P., and  Brinkmann, W. 2004 A\&A 422, 85.
   
\bibitem[\protect\citeauthoryear{}{}]{} 
Pottschmidt K., et al., 2003, A\&A, 407, 1039



\bibitem[\protect\citeauthoryear{}{}]{} 
Rappaport S.A., Podsiadlowski Ph., Pfahl E., 2005, MNRAS, 356, 401

\bibitem[\protect\citeauthoryear{}{}]{} 
Remillard R. A., Muno M. P., McClintock J. E., Orosz J. A., 2002, in
{\it New Views on Microquasars}, ed. Ph. Durouchoux, Y. Fuchs, and
J. Rodriguez, Center for Space Physics: Kolkata (India), 49
(astro-ph/0208402)

\bibitem[\protect\citeauthoryear{}{}]{}
Roberts T.P., Warwick R.S., 2000, MNRAS, 315, 98.

\bibitem[\protect\citeauthoryear{}{}]{}
Roberts T.P., Warwick R.S., Ward M.J., Goad M.R., Jenkins L.P., 2005,
MNRAS, 357, 1363

\bibitem[\protect\citeauthoryear{}{}]{} 
Soria R., Cropper M., Pakull M., Mushotzky R., Wu K., 2005, MNRAS, 356, 12

\bibitem[\protect\citeauthoryear{}{}]{} 
Soria R., Motch C., Read A.M., Stevens I.R., 2004, A\&A, 423, 955

\bibitem[\protect\citeauthoryear{}{}]{} 
Stobbart A., Roberts T.P., Warwick R.S., 2004, MNRAS, 351, 1063

\bibitem[\protect\citeauthoryear{}{}]{} 
Titarchuk L., 1994, ApJ, 434, 570

\bibitem[\protect\citeauthoryear{}{}]{} 
Uttley P., M$^{\rm c}$Hardy I. M., 2005, in prep.

\bibitem[\protect\citeauthoryear{}{}]{}
Uttley P., McHardy I.M., Papadakis I.E., 2002, MNRAS, 332, 231.
			      
									      
\bibitem[\protect\citeauthoryear{}{}]{} 
Vaughan S., Iwasawa K., Fabian A.C., Hayashida K., 2005, MNRAS, 356, 524

									      
\bibitem[\protect\citeauthoryear{}{}]{} 
van der Klis M., 2005, in {\it Compact stellar X-ray sources}, ed.,
W. H. G. Lewin, M. van der Klis, Cambridge University Press
(Cambridge), in press (astro-ph/0410551)

\bibitem[\protect\citeauthoryear{}{}]{}
Walter, R. and  Fink, H.H. 1993, A\&A 274, 105.

\bibitem[\protect\citeauthoryear{}{}]{} 
Wilms J., Allen A., McCray R., 2000, ApJ, 542, 914

									      
\bibitem[\protect\citeauthoryear{}{}]{} 
Zdziarski A.A., Gierlinski M., Rao A.R., Vadawale S.V., Mikolajewska
J., 2005, MNRAS, 360, 825

\bibitem[\protect\citeauthoryear{}{}]{}
Zezas A.L., Georgantopoulos I., Ward M.J., 1999, MNRAS, 308, 302


\bibitem[\protect\citeauthoryear{}{}]{}
Zhang S.N., Cui W., Chen W., Yao Y., Zhang X., Sun X., Wu X.-B., Xu
H., 2000, Science, 287, 1239






\end{thebibliography}
\end{document}